\documentclass{article}
\usepackage{todonotes}
\usepackage{pifont}

\PassOptionsToPackage{numbers, compress}{natbib}

\usepackage{graphicx}
\usepackage{subcaption}
\usepackage{xspace}
\usepackage{amsmath,amssymb}
\usepackage{float} 
\usepackage[table]{xcolor}
\usepackage{multirow}
\usepackage{tabularx,booktabs}
\usepackage[final]{neurips_2025_ml4ps}

\usepackage[utf8]{inputenc} 
\usepackage[T1]{fontenc}    
\usepackage{hyperref}       
\usepackage{url}            
\usepackage{booktabs}       
\usepackage{amsfonts}       
\usepackage{nicefrac}       
\usepackage{microtype}      
\usepackage{xcolor}         

\newcommand{\kt}{\ensuremath{k_{\mathrm{T}}}\xspace}
\newcommand{\pt}{\ensuremath{p_{\mathrm{T}}}\xspace}

\newcommand{\jetclass}{{\textsc{JetClass}}\xspace}
\newcommand{\pairembed}{{\textsc{PairEmbed}}\xspace}
\newcommand{\rej}[1]{\ensuremath{\text{Rej}_{#1}}\xspace}

\newcommand{\hbb}{\ensuremath{H\to b \bar{b}}\xspace}
\newcommand{\hcc}{\ensuremath{H\to c \bar{c}}\xspace}
\newcommand{\hgg}{\ensuremath{H\to g g}\xspace}

\newcommand{\hqqqq}{\ensuremath{H\to 4 q}\xspace}
\newcommand{\hlvqq}{\ensuremath{H\to \ell \nu q q'}\xspace}
\newcommand{\tbqq}{\ensuremath{t\to b q q'}\xspace}
\newcommand{\tblv}{\ensuremath{t\to b \ell \nu}\xspace}
\newcommand{\wqq}{\ensuremath{W\to q q'}\xspace}
\newcommand{\zqq}{\ensuremath{Z\to q \bar{q}}\xspace}
\newcommand{\qgj}{\ensuremath{q/g}\xspace}

\title{Why is Attention Sparse in Particle Transformer?}

\author{%
Timothy Legge$^{2}$  \thanks{Corresponding Author: tlegge@ucsd.edu}\And
Aaron Wang$^{1}$  \thanks{Corresponding Author: aaronw5@uic.edu}\And
Jacob Ortiz$^{2}$ \And
Victor Limouzi$^{2}$ \And
Zihan Zhao$^{2}$ \And
Abhijith Gandrakota$^{3}$ \And
Elham E. Khoda$^{2}$ \And
Jennifer Ngadiuba$^{3}$ \And
Javier Duarte$^{2}$ \And
Richard Cavanaugh$^{1}$ \\
\\
$^{1}$University of Illinois Chicago, Chicago, IL 60607 \\
$^{2}$University of California San Diego, La Jolla, CA 92093 \\
$^{3}$Fermi National Accelerator Laboratory, Batavia, IL 60510 \\
}

\begin{document}

\begin{flushright}
    FERMILAB-PUB-25-0820-CMS-LDRD-PPD
\end{flushright}

\maketitle

\begin{abstract}
Transformer-based models have achieved state-of-the-art performance in jet tagging at the CERN Large Hadron Collider (LHC), with the Particle Transformer (ParT) representing a leading example of such models. A striking feature of ParT is its sparse, nearly binary attention structure, raising questions about the origin of this behavior and whether it encodes physically meaningful correlations. In this work, we investigate the source of ParT’s sparse attention by comparing models trained on multiple benchmark datasets and examine the relative contributions of the attention term and the physics-inspired interaction matrix before softmax. We find that binary sparsity arises primarily from the attention mechanism itself, with the interaction matrix playing a secondary role. Moreover, we show that ParT is able to identify key jet substructure features, such as leptons in semileptonic top decays, even without explicit particle identification inputs. These results provide new insight into the interpretability of transformer-based jet taggers and clarify the conditions under which sparse attention patterns emerge in ParT. Our code is available \href{https://github.com/aaronw5/Interpreting-Particle-Transformers}{here}.

\end{abstract}
\section{Introduction}

Machine learning (ML) has become essential for analyzing the massive datasets from the CERN Large Hadron Collider (LHC)~\cite{Harris:2022qtm}. Jet tagging, or the classification of collimated streams of particles that result from high-energy collisions at LHC, has proven to be a task highly suitable to ML. Among the most promising approaches to jet tagging are graph neural networks (GNNs) and transformers, which capture correlations among particles in jets. GNNs, such as ParticleNet~\cite{Qu:2019gqs}, reached state-of-the-art jet tagging by treating jets as unordered particle clouds~\cite{Shlomi:2020gdn,Duarte:2020ngm}. More recently, transformer models~\cite{NIPS2017_3f5ee243}, which have transformed language and vision processing~\cite{ramesh2022hierarchicaltextconditionalimagegeneration,geminiteam2024geminifamilyhighlycapable,NEURIPS2020_1457c0d6}, have been adapted to collider physics, where their ability to model long-range dependencies provides a powerful tool for identifying jet substructure signatures~\cite{CMS-PAS-HIG-23-012,CMS-PAS-JME-25-001,CMS-PAS-HIG-24-008,CMS-PAS-HIG-24-010}.

A leading example is the Particle Transformer (ParT)~\cite{Qu2022}, which augments multi-head attention with physics informed pairwise particle interaction features. This has made ParT one of the most effective jet taggers to date. 

Interpretability is crucial in high-energy physics to ensure ML models capture physically meaningful correlations rather than dataset-specific artifacts. Attention provides a natural interpretability handle: particle--particle attention maps can reveal what transformers learn~\cite{Chefer_2021_CVPR}. Understanding when and why attention becomes sparse may further guide model optimization, clarifying the role of physics-inspired interactions and suggesting opportunities to simplify architectures and make them more efficient.

\section{Related Work}

ML methods, particularly graph neural networks (GNNs), have been widely applied to particle physics problems at the CERN LHC~\cite{Shlomi:2020gdn,Duarte:2020ngm}. ParticleNet~\cite{Qu:2019gqs}, a GNN designed for jet tagging, achieved state-of-the-art performance by treating jets as particle clouds. Recent studies have also interpreted such models. For example, Mokhtar \emph{et al.}~\cite{mokhtar2022graphneuralnetworkslearn} employed layerwise relevance propagation to construct edge relevancy, uncovering how ParticleNet distinguishes three-prong hadronic top quark decays. These efforts highlight the growing emphasis on explainability in ML for high-energy physics.

Building on this line of work, attention-based transformers have emerged as powerful alternatives to GNNs for jet tagging. ParT~\cite{Qu_2020,Wang:2024rup} extends the standard attention mechanism with physics-inspired pairwise particle interactions. While delivering excellent performance, ParT also raises important interpretability questions. Wang \emph{et al.}~\cite{Wang:2024rup} analyzed ParT’s attention maps and particle-pair correlations in the $\eta$--$\phi$ plane, revealing a sparse, nearly binary attention pattern where each particle focuses on at most one other particle. Their subjet-level studies further showed that ParT attends to leptons in semileptonic top decays and captures both intra- and inter-subjet structures in hadronic top and Higgs decays, suggesting that the model learns physically meaningful jet substructure.

In line with these efforts, our work investigates the origin of ParT's sparse attention by systematically comparing datasets and quantifying the roles of traditional attention and the physics-inspired interaction matrix.

\begin{figure}[t]
    \centering
    \begin{subfigure}[t]{0.3\textwidth}
        \centering
        \includegraphics[width=\linewidth]{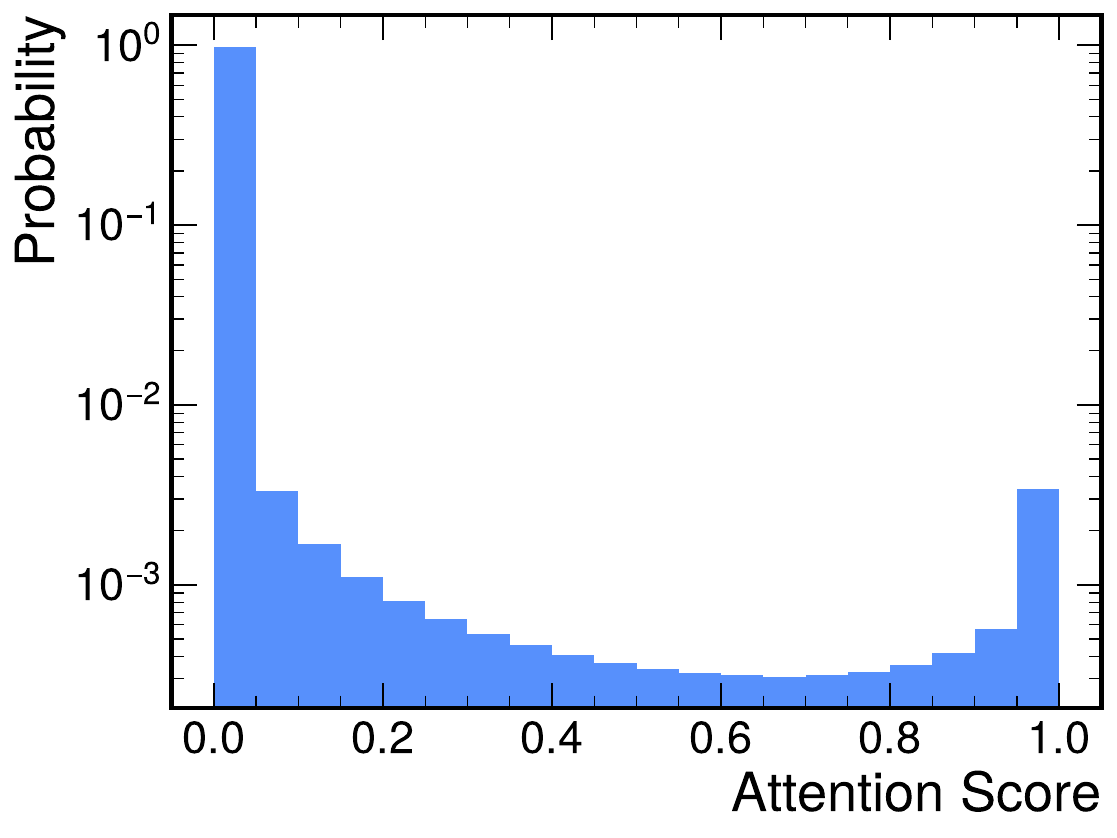}
        \caption{Quark--Gluon Dataset Attention Distribution}
        \label{fig:qgattn}
    \end{subfigure}
    \hfill
    \begin{subfigure}[t]{0.3\textwidth}
        \centering
        \includegraphics[width=\linewidth]{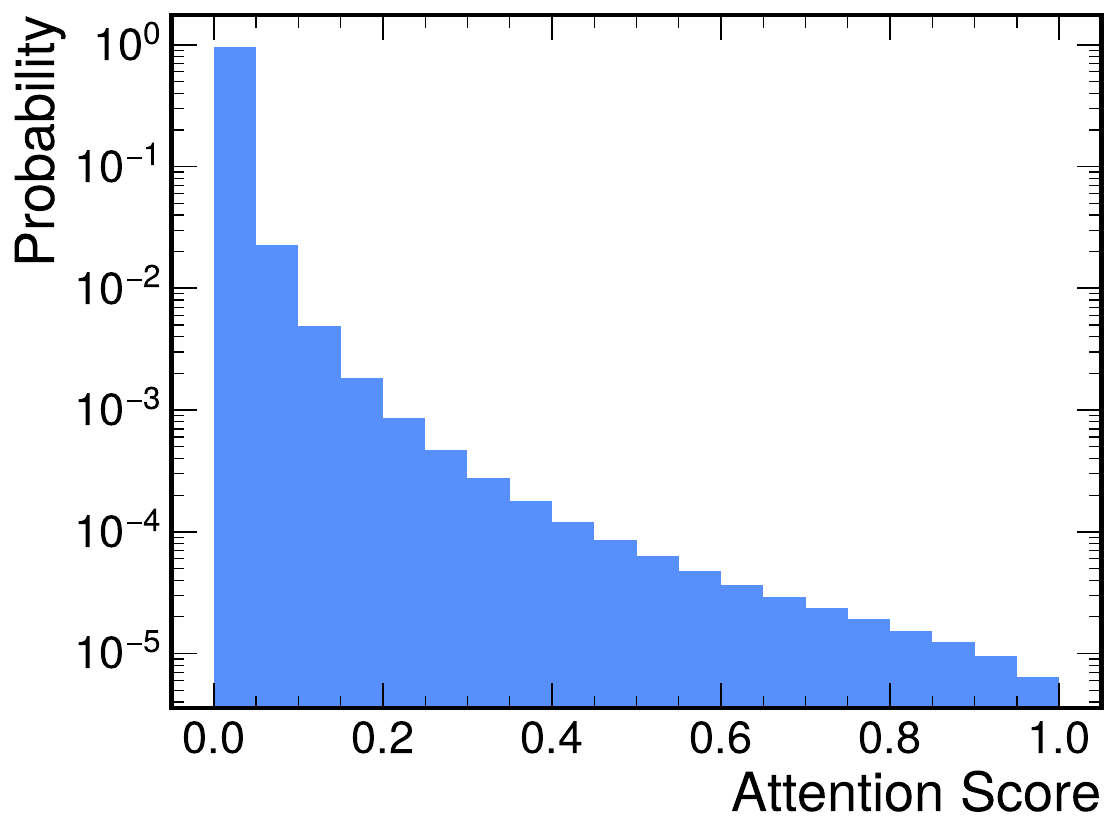}
        \caption{Top Landscape \tbqq Attention Distribution}
        \label{fig:toplandscape}
    \end{subfigure}
    \hfill
    \begin{subfigure}[t]{0.3\textwidth}
        \centering
        \includegraphics[width=\linewidth]{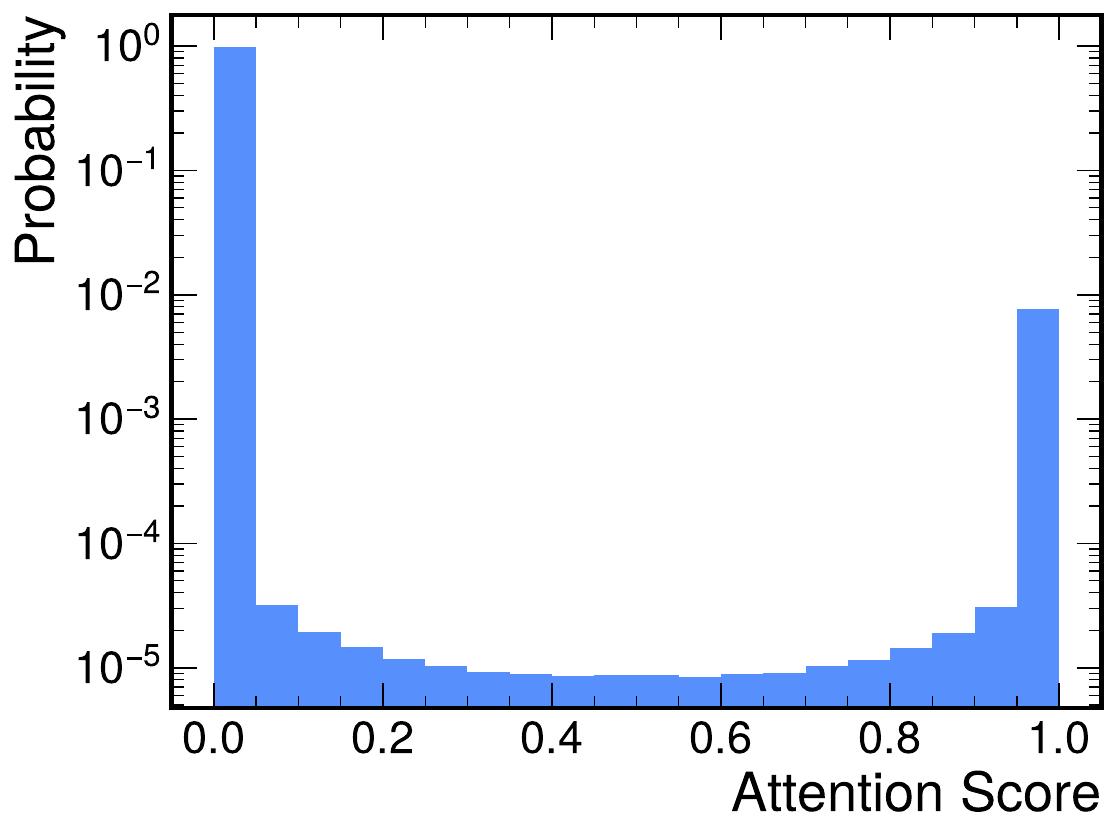}
        \caption{JetClass Full \tbqq Attention Distribution}
        \label{fig:jetclasstbqq}
    \end{subfigure}

    \vspace{0.5em}
    
    \begin{subfigure}[t]{0.3\textwidth}
        \centering
        \includegraphics[width=\linewidth]{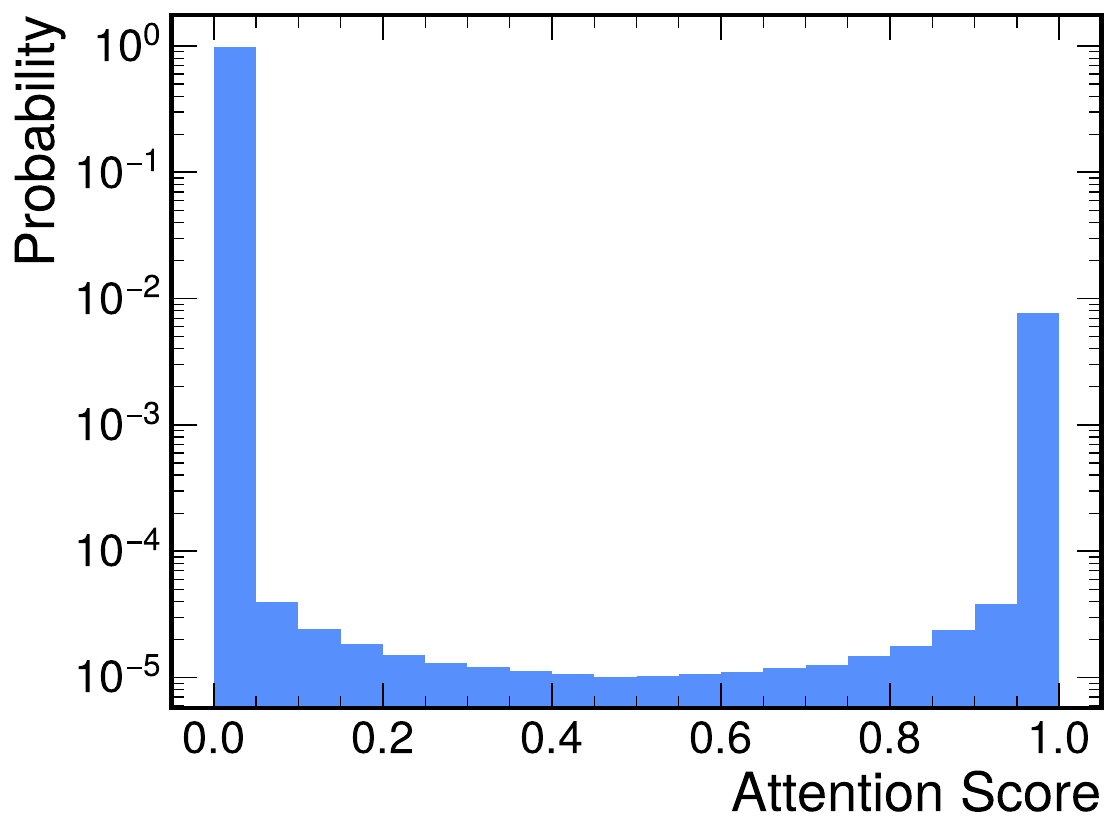}
        \caption{JetClass Kinematic \tbqq Attention Distribution}
        \label{fig:jetclasskintbqq}
    \end{subfigure}
    \hfill
    \begin{subfigure}[t]{0.3\textwidth}
        \centering
        \includegraphics[width=\linewidth]{Plots/JetClass_Full_Jet_0_Decay_TopHadronic_Layer_8_head_7.pdf}
        \caption{JetClass \tbqq}
        \label{fig:jctbqq}
    \end{subfigure}
    \hfill
    \begin{subfigure}[t]{0.3\textwidth}
        \centering
        \includegraphics[width=\linewidth]{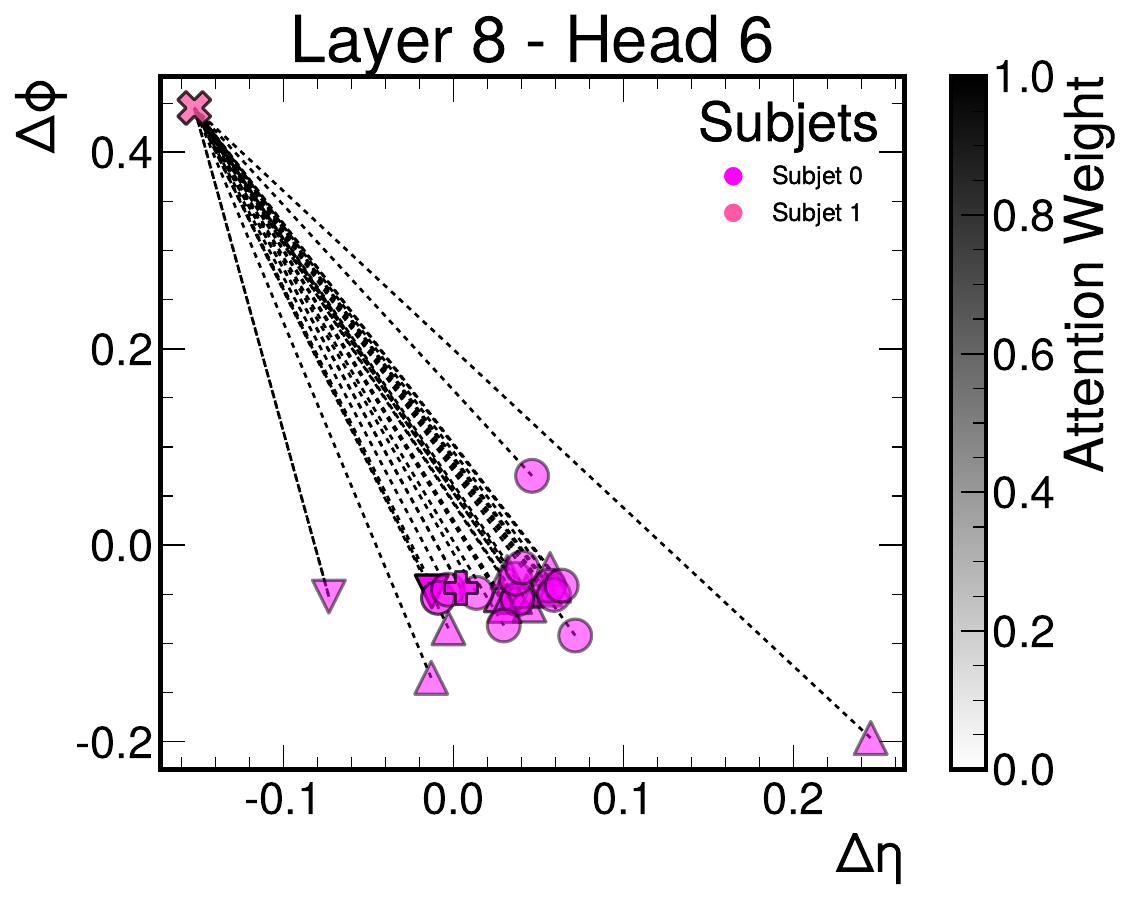}
        \caption{JetClass \tblv}
        \label{fig:jctbl}
    \end{subfigure}

    \vspace{0.5em}
    
    \begin{subfigure}[t]{0.3\textwidth}
        \centering
            \includegraphics[width=\linewidth]{Plots/JetClass_Kin_Jet_0_Decay_TopHadronic_Layer_8_head_1.pdf}
        \caption{JetClass Kinematic \tbqq}
        \label{fig:jcktbqq}
    \end{subfigure}
    \hfill
    \begin{subfigure}[t]{0.3\textwidth}
        \centering
        \includegraphics[width=\linewidth]{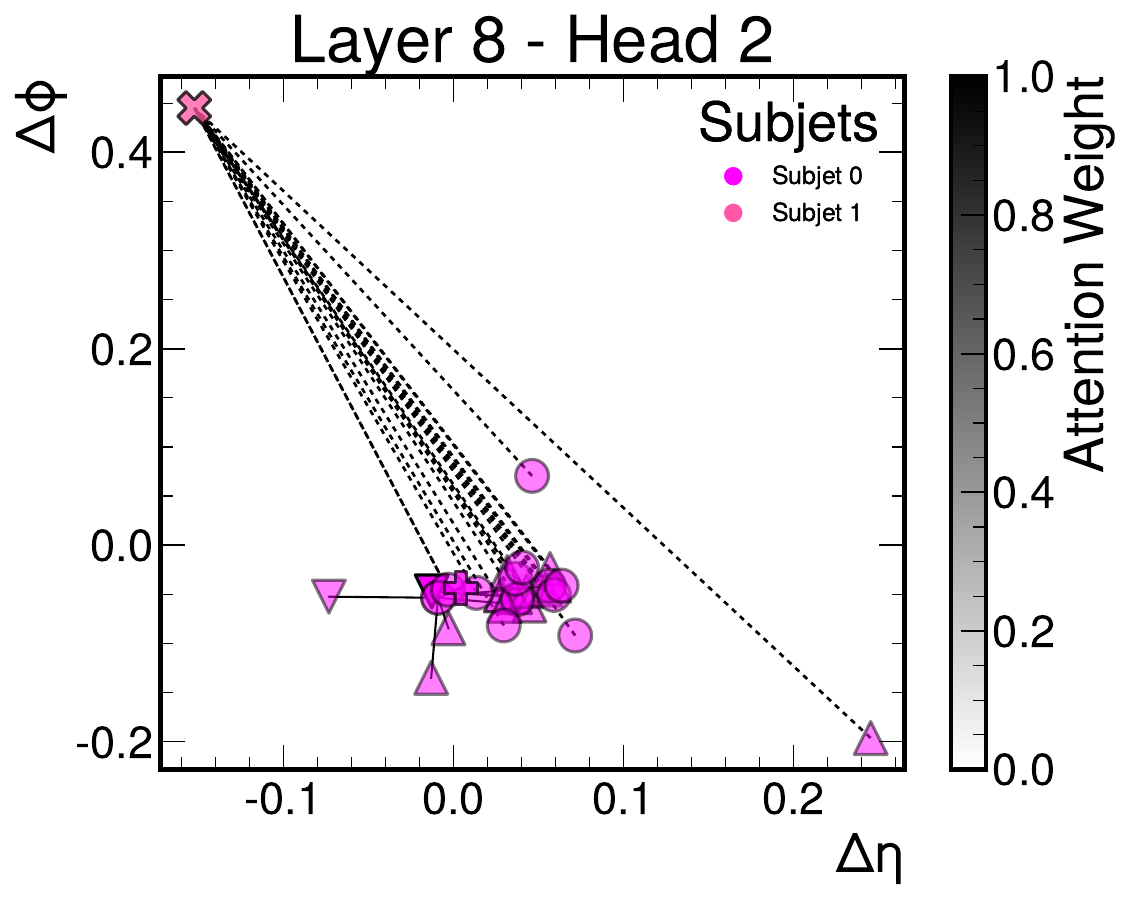}
        \caption{JetClass Kinematic \tblv}
        \label{fig:jcktbl}
    \end{subfigure}
    \hfill
    \begin{subfigure}[t]{0.3\textwidth}
        \centering
        \includegraphics[width=\linewidth]{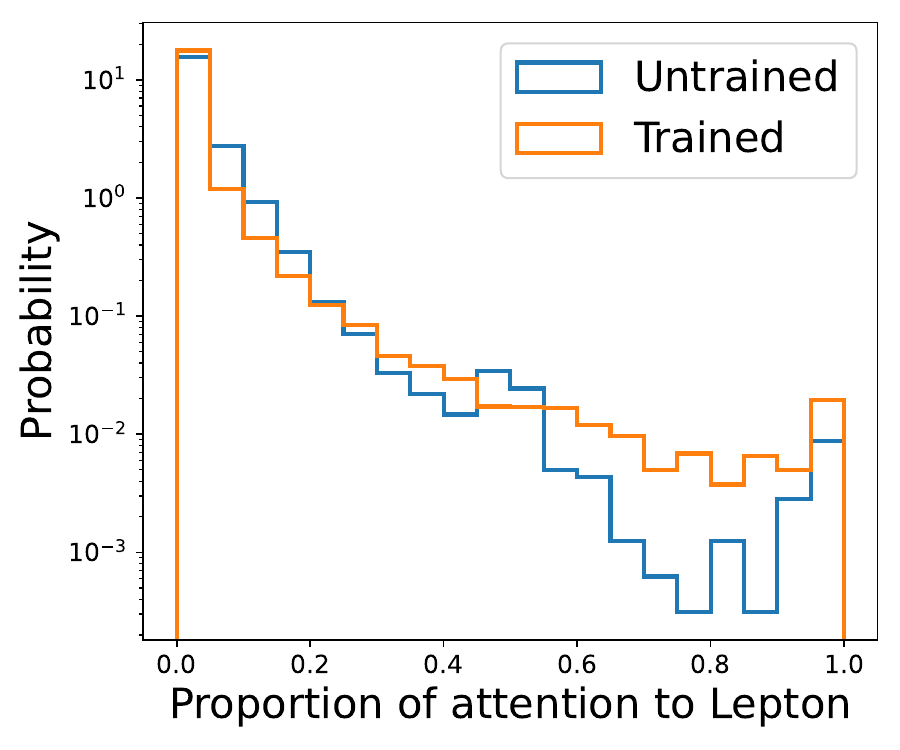}
        \caption{JetClass Kinematic Proportion of Attention to Lepton in \tblv }
        \label{fig:leptonAttention}
    \end{subfigure}
    
    \caption{Comparison of ParT attention distributions across datasets (a--d), $\eta$--$\phi$ particle attention maps (e--h), and the distribution of attention scores to the lepton in \tblv with ParT trained on JetClass Kinematic (i). $\eta$--$\phi$ maps use pre-softmax attention values without the interaction matrix included. Symbols: (\ding{54}: muon, \ding{115}: charged hadron, \ding{116}: neutral hadron, \ding{108}: photon, \ding{58}: electron). Transparency scales with particle $\pt$, while line intensity reflects attention scores.}
    \label{fig:attn_grid}
\end{figure}

\section{Datasets and Model}
This method compares the attention heatmaps and interpretability of multiple different datasets. We use the  \jetclass dataset~\cite{Qu2022}, the Top Landscape dataset \cite{kasieczka}, and the Quark--Gluon dataset as our benchmark datasets for comparative studies. 
\paragraph{Datasets} We evaluate models on three widely used jet-tagging benchmarks. The \jetclass dataset~\cite{jetclass} comprises 100M training, 5M validation, and 20M test jets across 10 classes from quarks, gluons, and decays of $W$, $Z$, Higgs, and top quarks (\qgj, \hbb, \hcc, \hgg, \hqqqq, \hlvqq, \tbqq, \tblv, \wqq, \zqq). Each jet contains up to 128 particles (avg. 30–50) with 17 features spanning kinematics, PID, and displacement. The Top Landscape Tagging benchmark~\cite{Kasieczka2019TopQuark} provides 2M jets (1.2M train, 0.4M val/test) distinguishing $t \to bqq'$ decays from light-quark/gluon backgrounds, described only by constituent four-vectors with 7 kinematic features.
Finally, the Quark--Gluon Tagging benchmark~\cite{komiske_2019_3164691} consists of 2M jets (1.6M train, 0.2M val/test) for discriminating quark- from gluon-initiated jets, using both kinematic and PID information.

\paragraph{Model Architecture} ParT is a state-of-the-art jet tagging model, which introduces a modified attention mechanism, called particle multi-head attention (P-MHA).  
Given a particle-level representation of a jet with $N$ constituents, $x \in \mathbb R^{N\times d}$, the attention is computed as
\begin{align}
\begin{split}
 \text{P-MHA}(x) &= \text{concat}(\text{head}_1, \dots, \text{head}_h)W^O \\
 \text{head}_i  &=  \text{softmax}\left(\frac{xW_i^W(xW_i^K)^{\intercal}}{\sqrt{d_k}} + U_i\right)xW_i^V,
 \label{eq:pmha}
 \end{split}
\end{align}
where $U_i\in \mathbb{R}^{N\times N}$ encodes pairwise particle interactions. These are learned from a set of kinematic features $(\ln \Delta, \ln k_{\text{T}}, \ln z, \ln m^2)$ described in  Refs.~\cite{Dreyer:2020brq,Qu2022} that are encoded through convolutional layers. We use pretrained ParT weights from Ref.~\cite{Qu2022} for models on the \jetclass dataset, and train from scratch on the Top and Quark--Gluon datasets.

\section{Results}
\paragraph{Distribution of Attention Scores} 
We visualize the distribution of attention scores across all heads for models trained on the full \jetclass feature set and on \jetclass with only kinematic features, as shown in images (a -- d) of Figure~\ref{fig:attn_grid}. For both cases, as well as for the Quark--Gluon dataset, ParT produces a very sparse, nearly binary attention map. In contrast, the Top Landscape dataset does not exhibit this binary behavior. Since the Top Landscape dataset contains only kinematic information and no particle identification (PID) features, we perform a direct comparison with the hadronic top (\tbqq) class in \jetclass using kinematic features only. We find that the attention distribution in \jetclass remains binary even without PID inputs, demonstrating that the binary nature of attention, as reported in previous work, does not rely on the presence of PID features~\cite{Wang:2024rup}.
\begin{figure}[t]
    \centering
    \begin{subfigure}[t]{0.32\textwidth}
        \centering
        \includegraphics[width=\linewidth]{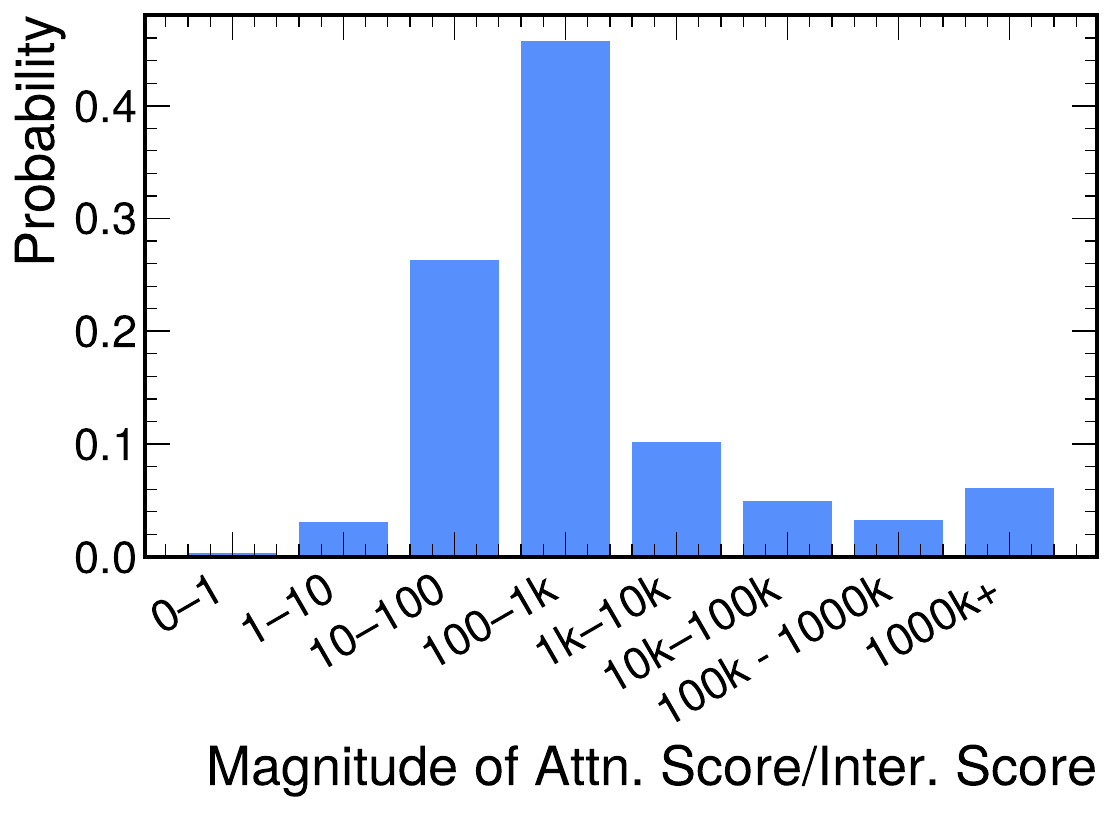}
        \caption{JetClass}
        \label{fig:jc_attnbar}
    \end{subfigure}
    \hfill
    \begin{subfigure}[t]{0.32\textwidth}
        \centering
        \includegraphics[width=\linewidth]{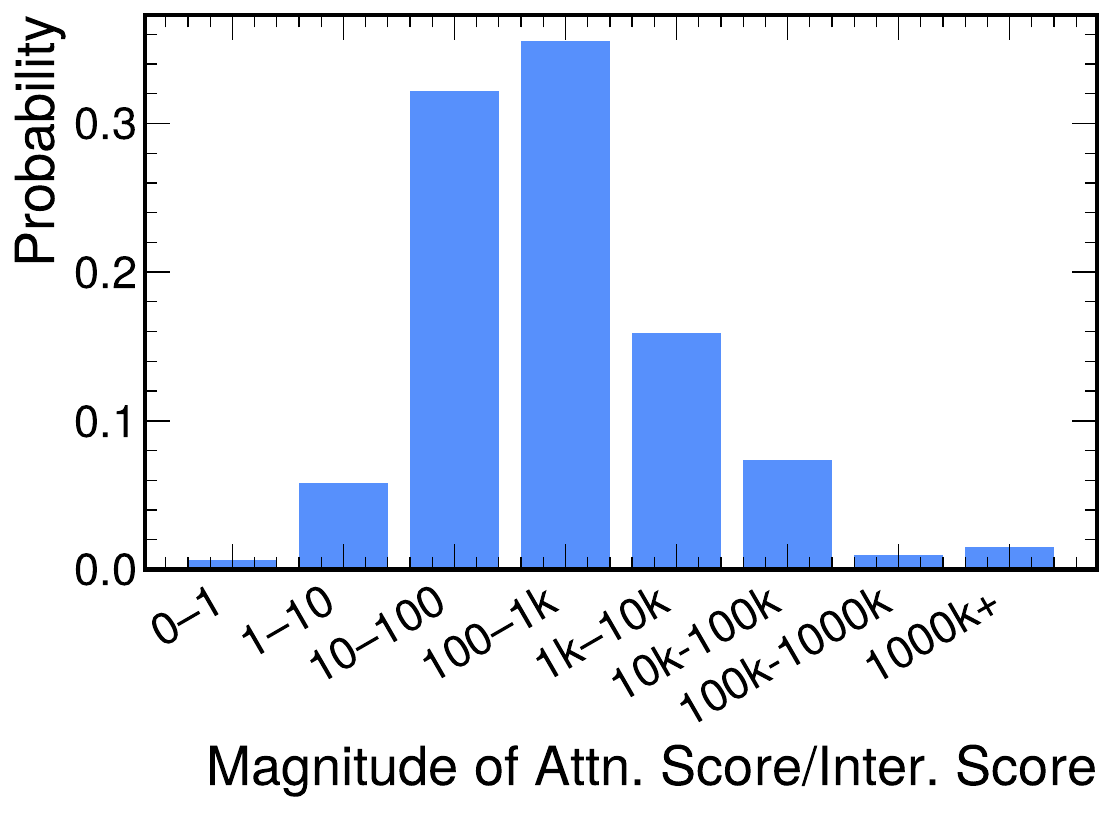}
        \caption{JetClass Kinematic}
        \label{fig:jck_attnbar}
    \end{subfigure}
    \hfill
    \begin{subfigure}[t]{0.32\textwidth}
        \centering
        \includegraphics[width=\linewidth]{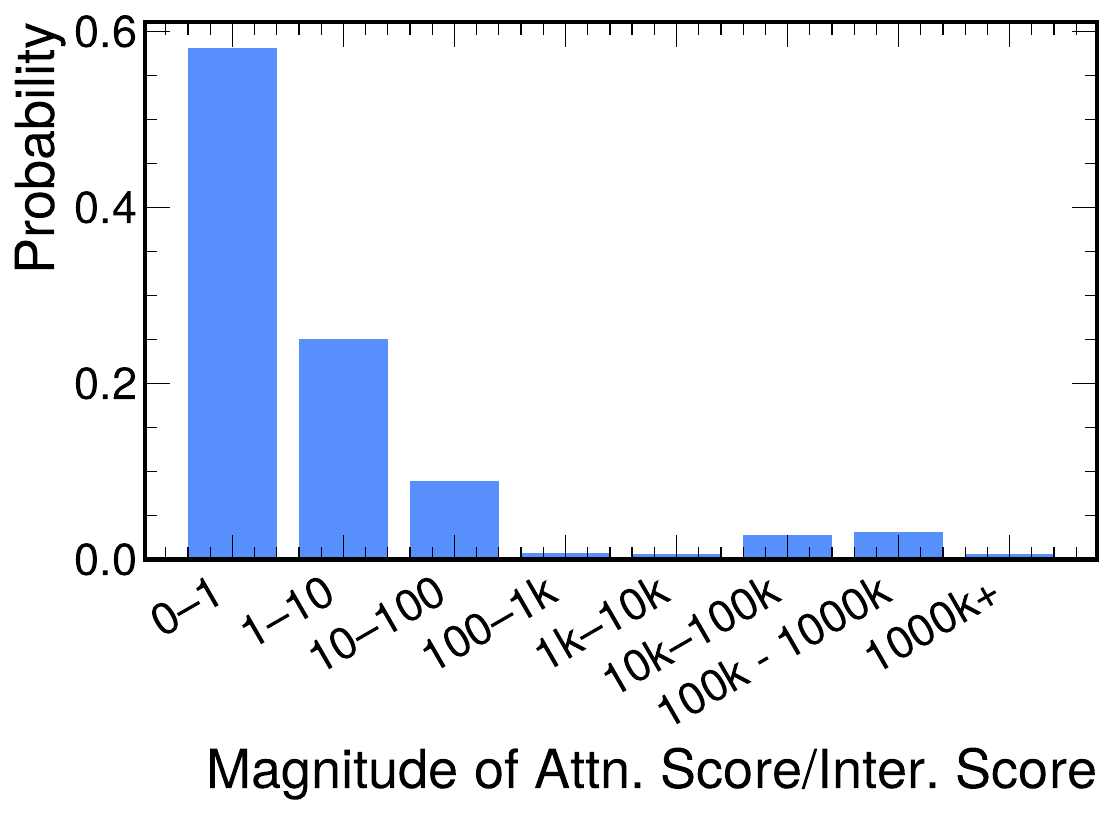}
        \caption{Top Landscape}
        \label{fig:tl_attnbar}
    \end{subfigure}
    
    \caption{Distribution of the magnitude of pre-softmax attention value divided by pre-softmax interaction matrix value. For both JetClass models, the ratio is almost always greater than one, and often $10^4$--$10^5$ times bigger than the interaction matrix, meaning it completely dominates.}
    \label{fig:attnBars}
\end{figure}

\paragraph{Interaction and Attention Score Distribution Before Softmax}
We compute the ratio between attention matrix values and their corresponding interaction matrix values, examining the resulting distribution in figure \ref{fig:attnBars}. For ParT trained on \jetclass, the attention term dominates the interaction matrix, indicating that it is the primary contributor to model performance. This shows that, even without the additional physics-motivated bias introduced by interaction, the attention mechanism alone is capable of identifying particles of interest, such as conventional jet substructure observables. In fact, despite benchmarks reported in~\cite{Qu2022}, our results suggest that the interaction matrix may not be essential for jet tagging. Given that the interaction matrix is computationally expensive, removing it could provide significant resource savings. By contrast, in the Top Landscape dataset, the interaction matrix and pre-softmax attention scores have comparable distributions, demonstrating that both terms contribute meaningfully to the overall attention.

\paragraph{Particle Attention Graphs} 
To further investigate the origin of sparsity, we decluster each jet into a fixed number of subjets using the \kt algorithm~\cite{Catani:1993hr} as implemented in the FastJet package~\cite{Cacciari:2011ma} with Python bindings, following the approach of Ref.~\cite{Wang:2024rup}. Specifically, we cluster into two subjets for leptonic top decays (\tblv), and three subjets for hadronic top decays (\tbqq). For each process, we visualize the attention maps by representing particles as points in the $\eta$--$\phi$ plane. Our analysis focuses on the attention heads in the final layer of P-MHA. Importantly, these visualizations are constructed using the \emph{pre-softmax attention values only}, isolating the contribution of the attention mechanism without the interaction matrix, and are compared with the corresponding maps in Ref.~\cite{Wang:2024rup}.

In Fig.~\ref{fig:attn_grid} (e--h), we find that even without the interaction matrix, the model is able to recover the same jet substructure observables. In addition, Figs.~\ref{fig:jcktbqq} and~\ref{fig:jcktbl} demonstrate that the Particle Transformer does not require particle identification information to attend to the lepton in leptonic top decays (\tblv) and is able to find subjet connections without PID or interaction matrix. Models trained with and without PID inputs both successfully recognize the lepton and attend to it. This demonstrates that jet substructure information is already learned in pre-softmax attention, and that explicit particle identification is not necessary for ParT to identify leptons. In Fig.~\ref{fig:leptonAttention}, we demonstrate that even without PID, in \tblv decays, JetClass Kinematic similarily attends to the lepton more often than an untrained model, signifying that it recognizes the lepton as an important particle index.

\paragraph{Ablation Study}
The preceding findings, particularly the overwhelmingly larger magnitudes observed for traditional attention scores, suggest the possibility that the ParT model as trained may not require the interaction matrix simply because its influence becomes minuscule in softmax computation. To test this, we conduct an ablation study of ParT with interaction removed, referred to as ParT (Zero \(U\)), with results shown in Table~\ref{tab:ablation-table}. Specifically, we implement this by setting all parameters in the pretrained model's \pairembed module to 0 and subsequently running inference on the \jetclass test set. This approach to ablation differs from Ref.~\cite{Qu2022}, wherein the model itself is trained without interaction and achieves metrics similar to the full architecture.

We find that ParT (Zero \(U\)) exhibits drastically reduced performance than full ParT and the ablated ParT (plain) from Ref.~\cite{Qu2022} (whose accuracies are 0.861 and 0.849 respectively). Although this is not surprising from a surface-level ML perspective---as manually modifying a significant number of weights necessarily destroys learned information---in this case it is remarkable the degree to which performance is impacted, considering the relatively minuscule scale of interaction scores.

To explain the performance difference and assess the impact of interaction at inference time, we investigate instances where the interaction matrix has significant influence in the softmax computation. As a metric, we choose to count instances where adding the interaction scores $U$ to traditional attention $A$ alters the index of the highest-weighted particle going in to the softmax computation. In mathematical terms, the criterion is that $\text{argmax}(A_j + U_j)\ne\text{argmax}(A_j)$ where $j$ indexes over all particles/tokens in the jet. These instances are referred to as interaction-dependent computations.

As a supplement to the study of interaction-dependence, we also track incidence of non-binary attention computation, which here is quantified as an instance where the largest post-softmax weight is $<$ 0.8. We track both overall incidence (the proportion of token updates to be non-binary or interaction-dependent computations) as well as the proportion of tokens to encounter a non-binary or interaction-dependent computation in any P-MHA head during inference. In order to assess if interaction-dependent computations are correlated with non-binary computations, we sort their incidence by head and compute the Pearson correlation coefficient (PCC) between them. Our results are collected in Table ~ref{tab:nb-and-interaction-dependent-table}.

Our results indicate that the vast majority of tokens encounter interaction-dependent computation at some point within the model, meaning that most token embeddings are affected by interaction. This explains the weak performance observed in Table~\ref{tab:ablation-table}. Our supplementary study of non-binary computation demonstrates that it is a less common process (though 42.1\% of tokens still encounter it at some point) and that there is no strong correlation between the occurrence of interaction-dependent vs. non-binary computations. From these results, we conclude that while traditional attention is responsible for the highly binary character of post-softmax weights, the inverse is not true: the interaction features are not responsible for the trace non-binary computations.

\begin{table*}[tb]
\caption{Performance of ParT with zeroed-out interaction when run on \jetclass test set (20M jets). The \rej{X\%} metric is the inverse of FPR at the threshold where TPR is \(X\)\%.}
\label{tab:ablation-table}
\begin{center}
\vskip -0.15in
\resizebox*{1\textwidth}{!}{
\begin{tabular}{lccccccccccc}
\toprule
                & \multicolumn{2}{c}{All classes}   & \hbb      & \hcc      & \hgg      & \hqqqq    & \hlvqq    & \tbqq     & \tblv     & \wqq      & \zqq      \\
                            & Accuracy  & AUC       &\rej{50\%} &\rej{50\%} &\rej{50\%} &\rej{50\%} &\rej{99\%} &\rej{50\%}&\rej{99.5\%}&\rej{50\%} &\rej{50\%} \\
\midrule
ParT (Zero \(U\))                & 0.405     & 0.8974    & 15.0      & 8.81     & 19.9       & 5.53      & 3.03      & 79.5     & 2.69     & 25.6       & 11.8      \\
\bottomrule
\end{tabular}
}
\end{center}
\vskip -0.2in
\end{table*}

\begin{table*}[tb]
\caption{Prevalence of non-binary and interaction-dependent computations when \(U\) is included, as well as the PCC between them.}
\label{tab:nb-and-interaction-dependent-table}
\begin{center}
\vskip -0.15in
\resizebox*{1\textwidth}{!}{
\begin{tabular}{cccc}
\toprule
    Type of complication    & \% Incidence overall   & \% of tokens to encounter      & PCC   \\
\midrule
Interaction-Dependence                        &           3.6              &             85.4                  &  \multirow{2}{*}{0.229}             \\
Non-binary Computation                          &           0.88             &      42.1                              \\
\bottomrule
\end{tabular}
}
\end{center}
\vskip -0.2in
\end{table*}

\section{Summary and Outlook}

The sparse, nearly binary attention observed in ParT emerges primarily from the attention mechanism itself rather than from the physics-inspired interaction matrix. However, it is also shown that interaction contributes as-of-yet unidentified biases which are necessary for full performance. This result clarifies the role of inductive biases in ParT and strengthens confidence that its learned patterns capture genuine jet substructure. Beyond interpretability, understanding the origin of sparsity also points to opportunities for improving efficiency, for example by adopting top-$k$ attention mechanisms that reduce computational cost without degrading performance.

\paragraph{Limitations}
Our study focuses mainly on the distributions of attention weights and their relative magnitudes, providing only a partial view of how sparsity relates to underlying physics. As emphasized in prior work, attention maps should not be interpreted as complete explanations of model behavior~\cite{DBLP:journals/corr/abs-1902-10186}. Further analyses will be needed to connect attention patterns more directly to physical observables. In addition, our results are limited to a fixed set of benchmark datasets and only one size of model architecture. Studying a broader range of physics processes and experimental conditions would be necessary to fully assess the universality of sparse attention.

\paragraph{Future Work}
Having shown that the interaction matrix cannot be fully disabled at inference time without drastic performance impacts, future work may study the exact ways in which interaction influences the model, for example by searching for trends in the particles which interaction assigns high scores to. If meaningful findings are made in this line of investigation, they could lead to new implemented physics biases that replicate interaction's behavior but operate at a lesser computational cost. Broadly, further work in interpretability may motivate physics-informed training strategies that retain performance while enabling leaner architectures.

\begin{ack}
A.G. and J.N. are supported by the DOE Office of Science, Award No. DE-SC0023524, FermiForward Discovery Group, LLC under Contract No. 89243024CSC000002 with the U.S. Department of Energy, Office of Science, Office of High Energy Physics, LDRD L2024-066-1, Fermilab, DOE Office of Science, Office of High Energy Physics ``Designing efficient edge AI with physics phenomena'' Project (DE-FOA-0002705), DOE Office of Science, Office of Advanced Scientific Computing Research under the ``Real-time Data Reduction Codesign at the Extreme Edge for Science'' Project (DE-FOA-0002501).
J.D., Z.Z., and T.L. are supported by the Research Corporation for Science Advancement (RCSA) under grant \#CS-CSA-2023-109, Alfred P. Sloan Foundation under grant \#FG-2023-20452, DOE, Office of Science, Office of High Energy Physics Early Career Research program under Award No. DE-SC0021187, DOE, Office of Science, Office of High Energy Physics under Award No. DE-SC0009919, and the U.S. National Science Foundation (NSF) Harnessing the Data Revolution (HDR) Institute for Accelerating AI Algorithms for Data Driven Discovery (A3D3) under Cooperative Agreement PHY-2117997.
This work was performed using the National Research Platform Nautilus HyperCluster supported by NSF awards CNS-1730158, ACI-1540112, ACI-1541349, OAC-1826967, OAC-2112167, CNS-2100237, CNS-2120019, the University of California Office of the President, and the University of California San Diego's California Institute for Telecommunications and Information Technology/Qualcomm Institute.
\end{ack}

\bibliographystyle{cms_unsrt}
\bibliography{bibliography}
\end{document}